# High-speed electrically driven liquid-crystal compact optical skyrmion encoder

Yu-Ping Tang[1,#], Zhenyu Guo[2,#], Ze-Yu Wang[1], Le Yu [1], Long-Yang Wang [1], Nilo Mata-Cervera[2], Yu Wang[2], Ning Wang[1], Yijie Shen[2,3,*], Ling-Ling Ma[1,*], Yan-Qing Lu[1,*]

[1]National Laboratory of Solid State Microstructures, Key Laboratory of Intelligent Optical Sensing and Manipulation, College of Engineering and Applied Sciences, Nanjing University, Nanjing 210023, China.

[2]Centre for Disruptive Photonic Technologies, School of Physical and Mathematical Sciences, Nanyang Technological University, Singapore 637371, Singapore

[3]School of Electrical and Electronic Engineering, Nanyang Technological University, Singapore 639798, Singapore

[#]These authors contributed equally

*Corresponding authors: yijie.shen@ntu.edu.sg (Y. Shen); malingling@nju.edu.cn (L.-L.M.); yqlu@nju.edu.cn (Y.-Q.L.)

## Abstract

Optical skyrmions possess topological polarization textures that can maintain topological robustness under external perturbations, making them promising carriers for disturbance-resistant optical information transmission. However, existing optical skyrmion generation schemes mostly rely on static optical elements or fixed nanostructures, making high-speed dynamic switching of the topological state difficult. Here, we propose a high-speed switchable optical skyrmion encoder based on a compact liquid-crystal spin-orbit device. The device employs the in-plane orientation of liquid crystals to imprint a fixed Pancharatnam-Berry geometric phase, while an applied voltage rapidly tunes the liquid-crystal retardance, enabling reversible switching between skyrmion and vortex-dominated near-zero-topology states. Experimental results show that the device exhibits millisecond electrical response, with bidirectional response times of 1.76 ms and 0.72 ms, corresponding to an ideal cycling rate of approximately 403 Hz, making it the fastest switchable optical skyrmion generator to date. Furthermore, by exploiting this rapid topological refreshing capability, we demonstrate voltage-addressed image encoding and decoding, providing a new liquid-crystal device platform for high-speed, refreshable, and compact topological optical information processing and potentially robust optical communication.

**Keywords**: optical skyrmions; liquid crystals; spin-orbit photonics; voltage-addressed topological state encoding; geometric phase; modal-balance control; Stokes polarimetry.

## Introduction

Skyrmions are topological quasiparticle that have attracted broad interest across disciplines, ranging from particle physics and condensed-matter physics to photonics[1–7]. In optical systems, the rich degrees of freedom of light provide a versatile platform for constructing skyrmion mappings. By engineering Stokes parameters[8–11], electromagnetic field vectors[12,13], spin angular momentum[14–17], and momentum-space pseudospin vectors[18–20], optical skyrmions have been generated and manipulated in multidimensional photonic parameter spaces. Among these implementations, Stokes skyrmions are particularly attractive because the spatial distribution of polarization states can be directly mapped onto the Poincaré sphere, allowing the skyrmion number to serve as a measurable topological descriptor for robust optical encoding[21,22]. Such topological structures can maintain distinguishable information under perturbations such as turbulence[23–25], scattering, and noise, making them promising carriers for disturbance-resistant optical information transmission[25].

Despite these prospects, most existing optical skyrmion encoding schemes remain essentially static. Free-space interferometric systems and spatial-light-modulator-based approaches can flexibly synthesize skyrmionic polarization textures[26,27], but they are usually bulky and not well suited for compact high-speed modulation. Metasurfaces[28–30], metaoptics and integrated photonic platforms have greatly reduced the device footprint and enabled complex spin-orbit transformations in planar or on-chip structures[31–33]. However, the generated skyrmion state is often fixed by a predefined device structure, making high-speed topological switching a key challenge for practical information applications[34,35]. Liquid crystals offer a promising route to electrically switchable optical skyrmion generation by combining spatially patterned spin-orbit coupling with voltage tunability[36–38]. In a Pancharatnam-Berry spin-orbit device, the written liquid-crystal orientation defines a fixed geometric-phase channel, whereas the electrically controlled retardance tunes the modal balance between Gaussian and vortex components. This enables rapid switching of the output Stokes texture between skyrmion and non-skyrmion states without rewriting the spatial pattern. This capability makes the liquid-crystal platform particularly suitable for compact, electrically addressable, and dynamically refreshable optical skyrmion systems.

Here we implement this concept using a patterned liquid-crystal spin-orbit device, in which the written director field defines a fixed Pancharatnam-Berry phase channel and the applied voltage tunes the retardance. At 633 nm, voltage-controlled Gaussian-vortex modal balancing drives the reconstructed Stokes texture to rapidly switch between skyrmion and non-skyrmion states. The device exhibits millisecond-scale bidirectional electrical switching, with characteristic response times of $1.76$ ms and $0.72$ ms, corresponding to an ideal bidirectional cycling rate of approximately 403 Hz. To the best of our knowledge, this represents the fastest switchable optical skyrmion encoder reported to date[39–41]. Skyrmion generation is maintained over most operating voltages and remains stable for $10^4$ s under continuous bias, demonstrating robust voltage-addressed topological control. We further demonstrate the generation of a second-order skyrmion texture, and exploit the electrically refreshed topological contrast to realize encoding and decoding of a $16 \times 16$, $256$-bit image. These results pave the way for high-speed, disturbance-resistant optical information transmission based on dynamically refreshable skyrmion states.

## Results

### Voltage-addressed topological evolution

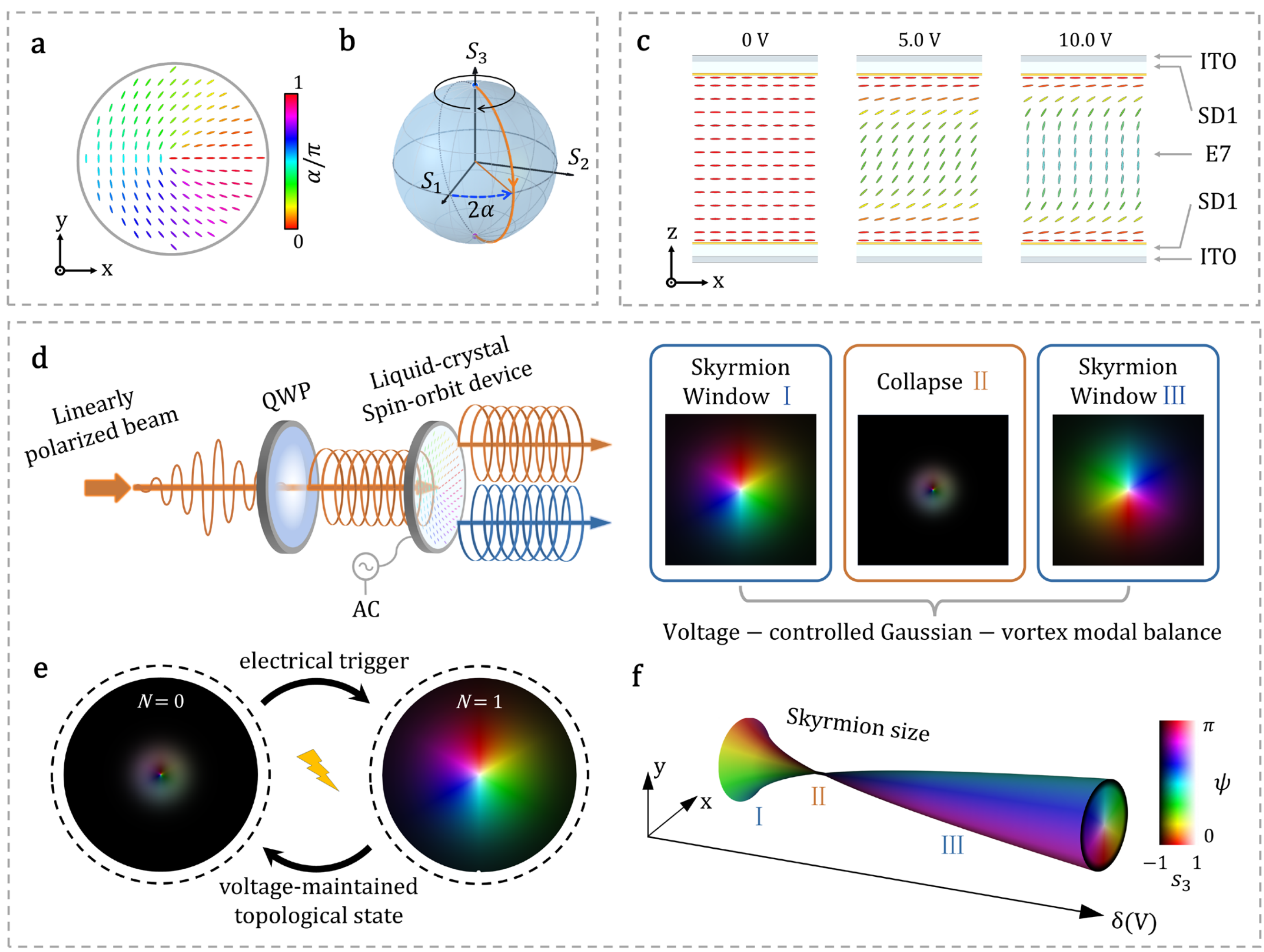

**Fig. 1 | Concept and control pathway of the voltage-addressed optical-skyrmion state encoder.**
**a** Patterned in-plane liquid-crystal director field $\alpha(x, y)$, which defines the fixed spatial Pancharatnam-Berry geometric-phase channel. **b** Geometric-phase-mediated spin-orbit conversion on the Poincaré sphere. The local optical-axis orientation introduces a spin-dependent phase $2\sigma\alpha(x, y)$ in the converted polarization channel. **c** Voltage-induced reorientation of the liquid-crystal directors. The applied alternating-current (AC) voltage voltage reduces the effective birefringence and tunes the retardance $\delta(V)$ without rewriting the in-plane director pattern. **d** Operating principle of the voltage-addressed optical-skyrmion generator. Voltage-controlled retardance changes the coherent balance between the Gaussian-like and vortex-like components, producing, within the selected branch surrounding the half-wave condition, two skyrmion windows separated by a vortex-dominated near-zero-topology collapse region. **e** Conceptual electrical addressing between a vortex-like near-zero-topology state and a skyrmion state under the corresponding applied bias. **f** Evolution of the effective skyrmion-density support as the Gaussian-vortex modal balance varies. The displayed radial extent represents the effective skyrmion size, rather than the intensity radius of the vortex-like mode.

Liquid crystals are self-organized anisotropic soft materials that combine spatially programmable molecular alignment with pronounced electro-optic responsiveness[42–44]. This dual capability allows a single patterned layer to encode a prescribed geometric-phase transformation while retaining dynamic control over the spin-orbit conversion. In the present device, these two degrees of freedom are assigned distinct roles. The patterned in-plane director field $\alpha(x, y)$ fixes the local optical-axis orientation and the associated Pancharatnam-Berry phase (Fig. 1a,b), thereby defining the spatial spin-orbit channel[45]. By contrast, the applied voltage induces director reorientation through the cell thickness and changes the effective birefringence and retardance without rewriting the in-plane pattern (Fig. 1c). Electrical addressing therefore controls the conversion amplitude while preserving the encoded spatial phase profile[46],

$$\delta(V,d,\lambda)=\frac{2\pi}{\lambda}\Delta n_{\mathrm{eff}}(V)d, \tag{1}$$

Here, $d$ is the cell gap, $\lambda$ is the operating wavelength, and $\Delta n_{\mathrm{eff}}(V)$ is the voltage-dependent effective birefringence. Equation (1) defines the electrical control coordinate: voltage changes the retardance without rewriting the patterned geometric-phase channel[42].

For the fixed incident circular polarization $|\sigma_0\rangle$, used in the experiments, the field at the selected observation plane can be written as

$$E_{\mathrm{out}}(r,\phi,V)=cos\left[\frac{\delta(V)}{2}\right]u_{\mathrm{G}}(r)|\sigma_0\rangle+i\,sin\left[\frac{\delta(V)}{2}\right]u_{\mathrm{L}}(r)e^{il\phi}|-\sigma_0\rangle, \tag{2}$$

Here, $u_{\mathrm{G}}(r)$ and $u_{\mathrm{L}}(r)e^{il\phi}$ are the propagated Gaussian-like and vortex-like field profiles at the observation plane, respectively, and $l$ is the azimuthal order. For the rotational pattern used here, the written geometric phase satisfies $2\sigma\alpha(x,y)=l\phi$, Equation (2) makes the control mechanism explicit: the patterned director field fixes the vortex phase[42,47,48], whereas $\delta(V)$ controls the modal weights through $a_{\mathrm{G}}(V)\propto cos\left[\frac{\delta(V)}{2}\right]$, $a_{\mathrm{L}}(V)\propto i\,sin\left[\frac{\delta(V)}{2}\right]$. This Gaussian-vortex redistribution forms the operating pathway in Fig. 1d and is measured through the two circular-polarization channels in Fig. 2. Defining $A_{\mathrm{G}}(r,V)=a_{\mathrm{G}}(V)u_{\mathrm{G}}(r)$ and $A_{\mathrm{L}}(r,\phi,V)=a_{\mathrm{L}}(V)u_{\mathrm{L}}(r)e^{il\phi}$, the normalized Stokes components in the ordered circular-polarization basis $(|\sigma_0\rangle,|-\sigma_0\rangle)$ are

$$s_3=\frac{|A_{\mathrm{G}}|^2-|A_{\mathrm{L}}|^2}{|A_{\mathrm{G}}|^2+|A_{\mathrm{L}}|^2},\qquad s_1+is_2=\frac{2A_{\mathrm{G}}^*A_{\mathrm{L}}}{|A_{\mathrm{G}}|^2+|A_{\mathrm{L}}|^2}. \tag{3}$$

The local modal ratio determines the radial variation of $s_3$, whereas the coherent cross term carries the azimuthal winding of $s_1$ and $s_2$. For $l=1$, their combination produces a first-order Stokes skyrmion when the two components jointly provide the radial pole-to-pole transition and the transverse winding.
Near the half-wave condition $\delta=\pi$, the Gaussian-like coefficient approaches zero, suppressing the transverse Stokes components and the complete Poincaré-sphere coverage. The output therefore approaches the vortex-like near-zero-topology state shown in (Fig. 1e). Away from this condition, the Gaussian-like component reappears and the skyrmion texture is restored.

Across the half-wave condition, the Gaussian-like coefficient changes sign. For equal retardance offsets $\Delta$,

$$s_\perp(\pi+\Delta)=-s_\perp(\pi-\Delta),\qquad s_3(\pi+\Delta)=s_3(\pi-\Delta), \tag{4}$$

where, $s_\perp(\pi+\Delta)=(s_1,s_2)$. Equation (4) predicts a $\pi$ rotation of the transverse Stokes texture about the $s_3$ axis, while retaining the principal $s_3$ distribution. The dominant skyrmion-density support is located near the modal-matching radius satisfying $|A_{\mathrm{G}}|\simeq|A_{\mathrm{L}}|$. Its radial extent contracts as the vortex-like component becomes dominant and re-expands when the Gaussian-vortex balance is restored, corresponding to the evolution of the effective skyrmion size in (Fig. 1f). Within the experimentally selected retardance branch surrounding the half-wave condition, the model therefore predicts a reversible $N_{\mathrm{sk}}\simeq 1\rightarrow 0\rightarrow 1$, trajectory, accompanied by contraction and recovery of the skyrmion-density support and sign reversal of $s_1$ and $s_2$. These predictions are tested through the voltage–retardance calibration and Stokes reconstruction in Figs. 2 and

3.

## Experimental validation of voltage-addressed optical-skyrmion switching

Guided by this picture, we first calibrated how the patterned liquid-crystal device converts voltage into spin-orbit modal redistribution. Crossed-polarizer microscopy revealed the characteristic rotational texture across the device aperture (Fig. 2a half-wave). Although the applied voltage continuously modified the colour and intensity of the observed pattern, the characteristic azimuthal symmetry of the written texture remained discernible throughout the voltage scan. These changes arise from the field-induced reduction of the effective birefringence of the E7 layer and the corresponding variation in optical retardance.[43]

The voltage-dependent spin-orbit conversion was resolved in the circular-polarization basis. The co-polarized channel contained the unconverted Gaussian-like component, whereas the spin-flipped channel carried the geometric-phase vortex-like component[37,38,42]. The normalized line profiles at 4.4, 5.0 and 7.6 V identify the two modal contributions and show a vortex-dominated output near 5.0 V (Fig. 2b). Their integrated energies further quantify the voltage-dependent redistribution between the two channels (Fig. 2c).

The corresponding retardance trajectory was obtained from the measured converted-channel fraction. In the ideal lossless-retarder approximation,

$$F_{-\sigma}(V) \approx sin^2\left[\frac{\delta(V)}{2}\right] \tag{5}$$

Because the conversion fraction is periodic in $\delta$, the physical retardance branch was determined from the continuity of the voltage scan, the independently estimated initial retardance, and the monotonic decrease in effective birefringence under increasing voltage. The inferred retardance decreased continuously with voltage and passed through the half-wave condition near the maximum of the vortex-like output (Fig. 2c)[42]. This voltage-retardance calibration was subsequently used to track the topological evolution of the reconstructed Stokes field.

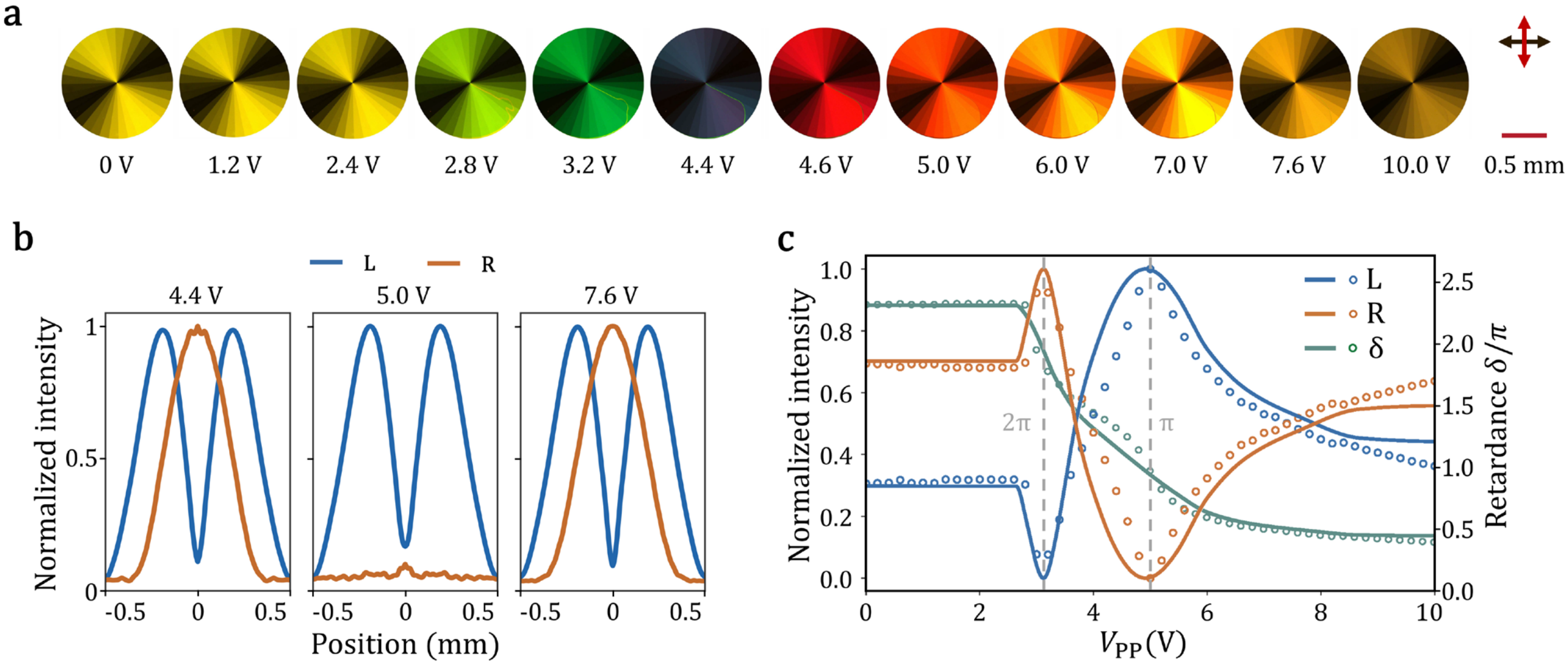

**Fig. 2 | Patterned director field and voltage-retardance calibration.**

**a** Cross-polarized micrographs of the patterned liquid-crystal spin-orbit device under different driving voltages, showing the written rotational director texture and its voltage-dependent birefringence response. Scale bar, $0.5$ mm. **b** Normalized centre-line profiles of the two circular-polarization channels at representative voltages, identifying the Gaussian-like and vortex-like modal contributions. **c** Integrated spin-channel energies and inferred retardance $\delta(V)$, showing the voltage-dependent modal redistribution. Open circles denote experimental data, solid lines are guides to the eye, and grey dashed lines mark the $\delta \approx 2\pi$ and $\delta \approx \pi$ coordinates.

With the voltage–retardance relation established, we next reconstructed the spatially resolved Stokes field to experimentally test the predicted window–collapse–window topological evolution. At 633 nm, we focused on the voltage interval spanning the half-wave condition, within which the reconstructed Stokes field underwent a reversible topological transition. The normalized Stokes vector was obtained from six polarization-resolved intensity measurements, and the skyrmion number was calculated over the adaptively extracted support region, as described in Methods. In skyrmion window I, $N_{\mathrm{sk}}$ remained close to unity. It then decreased sharply to nearly zero around $5.0$ V before recovering to a near-integer plateau in skyrmion window III (Fig. 3a,b). The voltage scan thus produced a clear $N_{\mathrm{sk}} \approx 1 \rightarrow 0 \rightarrow 1$ evolution.

The effective skyrmion size, defined by the radial extent of the reconstructed skyrmion-density support, changed concurrently with $N_{\mathrm{sk}}$. With increasing voltage, the effective skyrmion-density support progressively contracted towards the beam centre and reached its minimum extent near the collapse region. Beyond this voltage, the extended polarization texture reappeared and gradually expanded (Fig. 3a).

Representative measurements illustrate this evolution in greater detail (Fig. 3c). At 3.6 and 4.4 V, the textures exhibited a continuous azimuthal winding and yielded $N_{\mathrm{sk}} = 0.997$ and 0.987, respectively. Near 5.0 V, the texture was reduced to a weak localized structure and $N_{\mathrm{sk}}$ approached zero. At 5.4, 7.6 and 10.0 V, the extended skyrmion texture was progressively restored, with corresponding $N_{\mathrm{sk}}$ values of 0.915, 0.944 and 0.997.

The component-resolved Stokes maps reveal an additional symmetry across the collapse region (Fig. 3d). The spatial distributions of $s_1$ and $s_2$ on the two sides of the half-wave condition had opposite signs, whereas the characteristic core–ring distribution of $s_3$ was largely preserved. This sign reversal corresponds to a $\pi$ rotation of the transverse Stokes texture about the $s_3$ axis. Near 5.0 V, $s_1$ and $s_2$ were strongly suppressed, and the reconstructed field no longer covered the Poincaré sphere completely. At higher voltages, the transverse components reappeared with the opposite orientation, accompanied by the recovery of the skyrmion number. These observations are consistent with the voltage-controlled Gaussian-vortex modal-balance picture developed above.

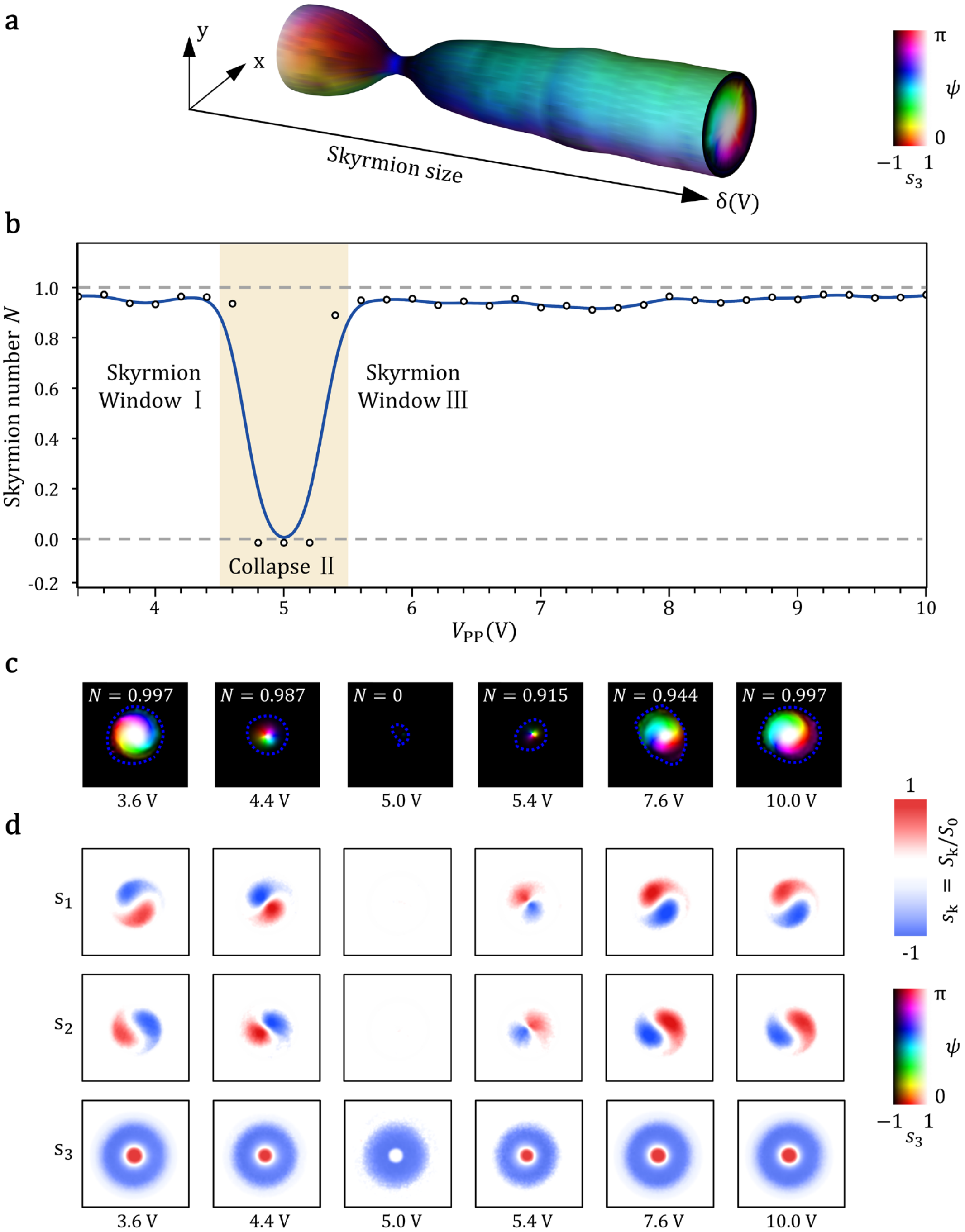

**Fig. 3 | Voltage-defined skyrmion windows and reversible topological collapse at 633 nm.**
**a** Experimental Stokes textures under different driving voltages, showing the contraction and re-expansion of the effective skyrmion support region. Hue denotes the local polarization azimuth $\psi = 1/2\, arg(s_1 + i\, s_2)$, and brightness denotes $s_3$. **b** Measured $N_{\mathrm{sk}}$ as a function of voltage. Points denote experimental values, the solid line is a visual guide, and the shaded region marks collapse region II between skyrmion windows I and III. **c** Representative Stokes textures at selected voltages, with measured $N_{\mathrm{sk}}$ values labelled. Blue dashed contours indicate the effective support region used for $\mathrm{N_{sk}}$ integration. **d** Normalized Stokes components $s_1$, $s_2$ and $s_3$ at the same voltages, showing sign reversal of $s_1$ and $s_2$ across the half-wave condition while the $s_3$ core-ring structure is largely retained.

Having established voltage-selective switching between skyrmion and near-zero-topology states, we next examined whether an electrically addressed skyrmion state can be held stably under continuous bias. Under continuous driving at 7.6 $V_{\mathrm{PP}}$ and 10 kHz, $N_{sk}$ remained on a stable near-unity plateau over $10^4$ s (Fig. 4a). The reconstructed textures recorded at 0, $10^2$ and $10^4$ s gave $N_{sk} = 0.955$, 0.957, and 0.956, respectively. The corresponding $s_1$, $s_2$ and $s_3$ maps were also nearly unchanged over the same interval, preserving both the transverse winding and the core-ring distribution of $s_3$ (Fig. 4b). These measurements show that the selected skyrmion state remains stable under continuous electrical driving.

Beyond the long-term holding of an addressed skyrmion state, we next examine whether the same retardance-controlled mechanism extends to a higher-order texture[34,35,38,39]. To this end, we use an independently fabricated second-order liquid-crystal spin-orbit device, which exhibits the same window-collapse-window evolution: two voltage windows with $N_{sk}$ close to $-2$ are separated by a near-zero-topology collapse region (Fig. 4c). In the two operating windows, the measured skyrmion number remains near $-2$, whereas around the collapse voltage the reconstructed texture approaches a near-zero-topology state. Representative textures at 4.0 and 8.8 V show the expected double azimuthal winding, with $N_{sk} = -1.925$ and $-1.961$, respectively, while the texture at 5.6 V collapses to near-zero topology (Fig. 4d). These results show that the voltage-addressed window-collapse-window response is retained for a higher-order skyrmion texture.

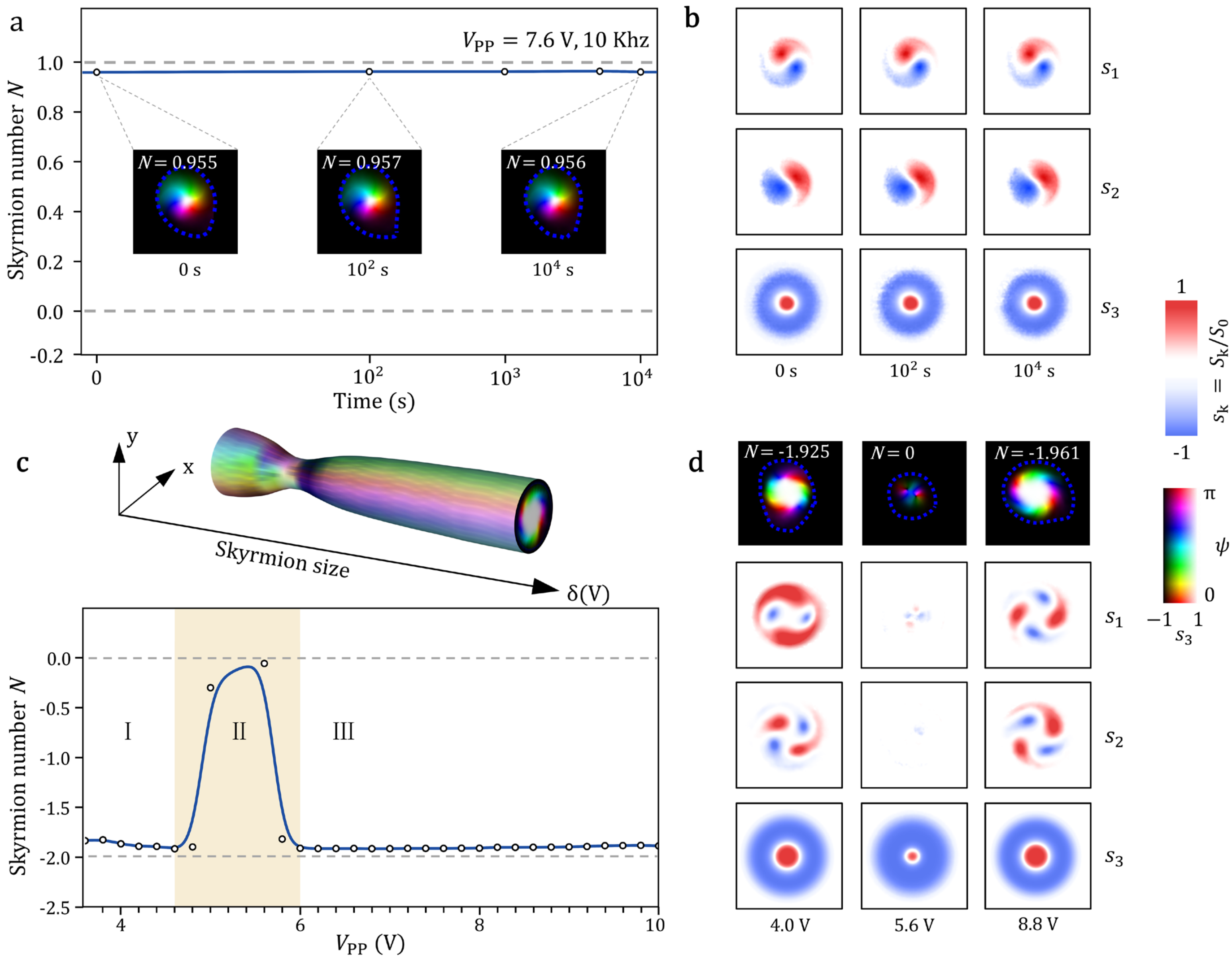

**Fig. 4 | Voltage-maintained skyrmion holding and second-order topological extension.**

**a** Long-time evolution of $N_{sk}$ under a fixed driving voltage. Insets show representative skyrmion textures at $t = 0$, $10^2$ and $10^4$ s, with the corresponding $N_{sk}$ values. **b** Normalized Stokes components $s_1$, $s_2$ and $s_3$ at the same times, showing preservation of the addressed texture under continuous bias. **c** Voltage response of a second-order optical skyrmion device, with $N_{sk}$ remaining close to $-2$ in skyrmion windows I and III and decreasing to near zero in collapse region II. **d**

Representative second-order Stokes textures and corresponding normalized Stokes components at selected voltages.

## Electrically refreshed topological-state encoding and image reconstruction

We next used the voltage-addressed vortex-skyrmion transition to demonstrate a dynamically refreshed topological-state information alphabet. A first-order device with a zero-voltage retardance close to the half-wave condition was selected for this experiment. At 0 V, the spin-orbit conversion was nearly complete and the output was dominated by the vortex-like component, producing a near-zero skyrmion number. At 4 V, the Gaussian-like contribution reappeared and formed a first-order skyrmion through coherent superposition with the vortex-like component. Polarization-resolved Stokes measurements confirmed the distinct vortex-like and skyrmion textures at the two voltage levels (Fig. 5a).

To determine the refresh rate of this topological alphabet, we measured the electrical switching dynamics between the calibrated logic states (Fig. 5b). The transient response was recorded through the normalized optical signal $R_{\mathrm{norm}}$ in the polarization channel providing the highest contrast between the two calibrated states. Repeated switching between 0 and 4 V produced reproducible transitions between two stable optical levels. The vortex-to-skyrmion transition reached its steady level within 0.72 ms, whereas the reverse skyrmion to vortex transition required 1.76 ms. These response times give an intrinsic bidirectional switching period of approximately 2.48 ms, corresponding to an upper cycling rate of about 403 Hz.

The clear optical contrast and millisecond switching speed then allowed the two voltage-addressed states to be used as binary topological symbols (Fig. 5c). The vortex-like state at 0 V represented logic 0, while the skyrmion state at 4 V represented logic 1. A $16 \times 16$ binary target image was rasterized into a 256-bit sequence and converted into the corresponding electrical waveform (Fig. 5d,e). During serial driving, $R_{\mathrm{norm}}$ followed the programmed voltage sequence between the two calibrated levels, retaining a clear optical contrast throughout the complete bit stream.

To verify that the written bits retained their topological identities rather than merely producing intensity contrast, we performed polarization-resolved Stokes reconstruction on representative states. Within a representative segment of the sequence, the measured $N_{\mathrm{sk}}$ values alternated between near-zero and near-unity levels in agreement with the programmed binary data (Fig. 5f). The corresponding reconstructed textures displayed the expected vortex-like and skyrmion polarization distributions (Fig. 5g). Across the full sequence, the measured states formed two well-separated $N_{\mathrm{sk}}$ populations associated with logic 0 and logic 1 (Fig. 5h).

The decoded topological states were finally rearranged according to the original raster order. The reconstructed $16 \times 16$ pattern reproduced the target image (Fig. 5i), completing the sequence from electrical waveform generation and topological-state writing to polarization-resolved readout and image recovery. These results demonstrate that the electrically refreshed vortex-skyrmion contrast can serve as a dynamically addressable topological information alphabet.

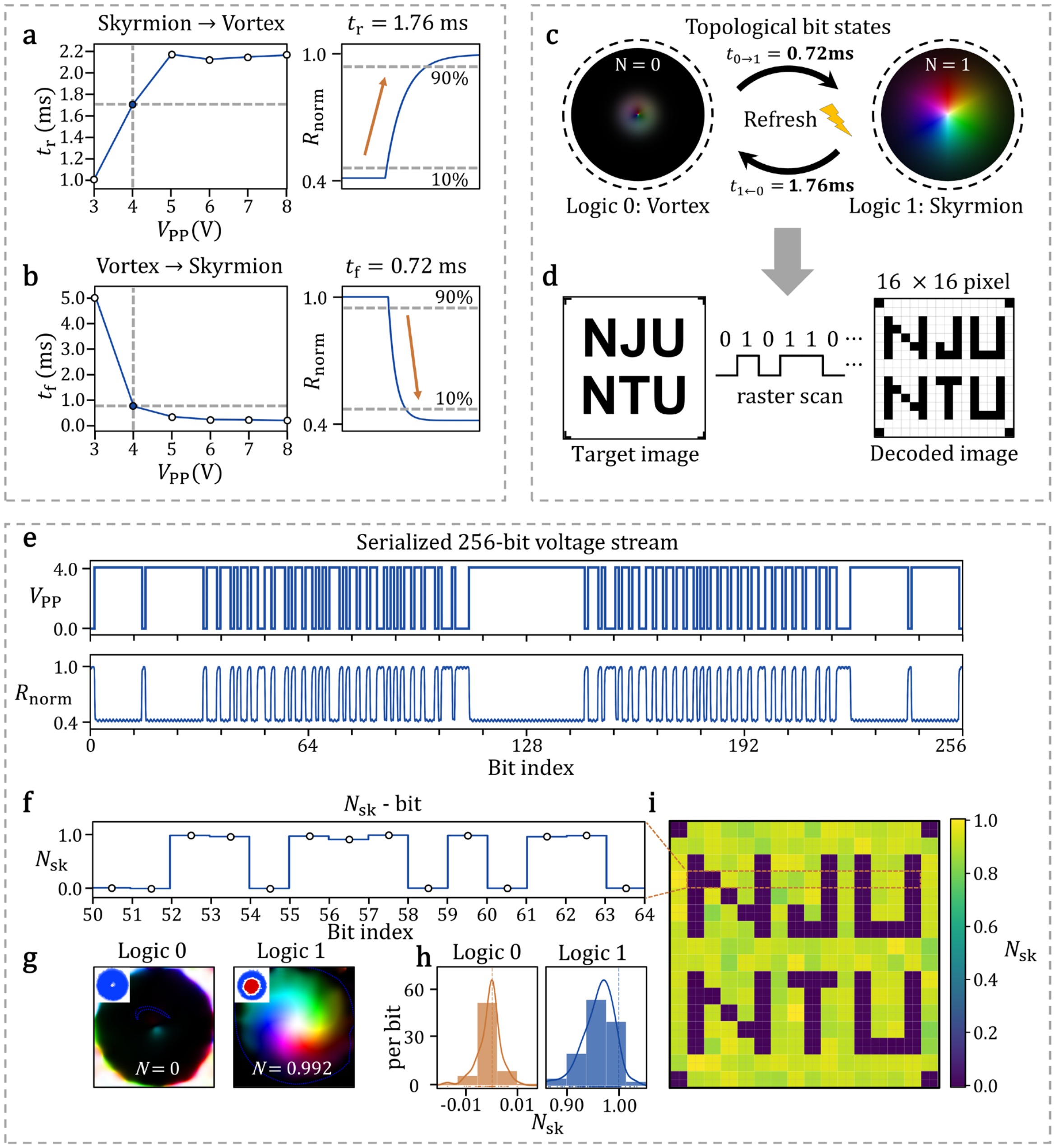

**Fig. 5 | Electrically refreshed topological bit encoding and 256-bit image reconstruction.**
**a,b** Bidirectional electrical switching between vortex-like and skyrmion states. The normalized readout signal $R_{\mathrm{norm}}$ reports the millisecond optical transient, with characteristic times of 1.76 ms for skyrmion to vortex switching and 0.72 ms for vortex to skyrmion switching. **c** Binary topological-state definition, assigning the vortex-like near-zero-topology state to logic 0 and the skyrmion state to logic 1. **d** Conversion of a $16 \times 16$ binary target image into a 256-bit raster-scanned sequence. **e** Serial $0/4$ V driving waveform and corresponding $R_{\mathrm{norm}}$ response. **f** $N_{\mathrm{sk}}$ readout over a representative local bit interval. **g** Representative reconstructed textures for logic 0 and logic 1.**h** Distribution of measured $N_{\mathrm{sk}}$ values for the two logic states, showing clear binary separation. **i** $16 \times 16$ image reconstructed from the measured $N_{\mathrm{sk}}$ values.

## Discussion

The central advance of this work is the separation of static spatial phase encoding from dynamic modal-balance control in a compact liquid-crystal spin-orbit device. This separation distinguishes the present platform from free-space, SLM-based[26,27], metasurface[28,33], and detuned-q-plate approaches[38], where either the system is bulky, the topology is largely fixed, or the tuning relies on changing the optical wavelength. The fixed in-plane director field preserves the Pancharatnam-Berry phase channel, whereas the electrically tuned retardance controls the coherent balance between Gaussian-like and vortex-like components. This design enables optical skyrmion states to be electrically refreshed without rewriting the spatial phase pattern, and reveals a voltage-addressed window-collapse-window pathway in which the Stokes texture reversibly switches between near-integer skyrmion states and a vortex-dominated near-zero-topology state. By exploiting this millisecond vortex-skyrmion contrast as a binary topological alphabet, we further demonstrate a complete 256-bit serial state-writing and polarization-resolved image-reconstruction workflow. These results move optical skyrmion generation from static-state preparation toward electrically refreshable topological-state modulation in a compact liquid-crystal platform. Future system-level implementation may require faster liquid-crystal materials, thinner cells, parallel Stokes readout, and array-addressed device architectures to close the gap between electrical writing, topological-state identification, and information throughput. In particular, emerging ferroelectric nematic liquid crystals provide a promising route toward much faster electro-optic response[49]. Integrating fast liquid-crystal materials with patterned spin-orbit devices may further extend the present platform from millisecond topological refreshing toward high-bandwidth topological photonic modulation[50–53]. This electrically tailored liquid-crystal topological photonic platform may open new opportunities for dynamic structured-light communication, reconfigurable optical information processing, and topology-enabled light-matter interaction[17,22,54–57].